# Vision Language Models versus Machine Learning Models Performance on Polyp Detection and Classification in Colonoscopy Images

## Running Title: Large Language Vision vs. Classical Machine Learning

Mohammad Amin Khalafi[1†], Seyed Amir Ahmad Safavi-Naini [1,2,3†], Ameneh Salehi[1], Nariman Naderi[1], Dorsa Alijanzadeh[1], Pardis Ketabi Moghadam[1], Kaveh Kavosi[4], Negar Golestani[2], Shabnam Shahrokh[1], Soltanali Fallah[5], Jamil S Samaan[6], Nicholas P. Tatonetti[7,8,9], Nicholas Hoerter[10], Girish Nadkarni[2,3], Hamid Asadzadeh Aghdaei[1*], Ali Soroush[2,3,10*]

1.     Research Institute for Gastroenterology and Liver Diseases, Shahid Beheshti University of Medical Sciences, Tehran, Iran
2.     Division of Data-Driven and Digital Health (D3M), The Charles Bronfman Institute for Personalized Medicine, Icahn School of Medicine at Mount Sinai, New York, NY, USA
3.     The Charles Bronfman Institute for Personalized Medicine, Icahn School of Medicine at Mount Sinai, New York, NY, USA
4.     Laboratory of Complex Biological Systems and Bioinformatics (CBB), Department of Bioinformatics, Institute of Biochemistry and Biophysics (IBB), University of Tehran, Tehran, Iran
5.     Department of GI Diseases, Tehran Milad Hospital, Tehran, Iran
6.     Karsh Division of Gastroenterology and Hepatology, Cedars-Sinai Medical Center, Los Angeles, CA
7.     Department of Computational Biomedicine, Cedars-Sinai Medical Center, West Hollywood, California, USA
8.     Cedars-Sinai Cancer, Cedars-Sinai Medical Center, 8700 Beverly Blvd. Los Angeles, CA, USA
9.     Department of Biomedical Informatics, Columbia University, New York, New York, USA.
10.    Henry D. Janowitz Division of Gastroenterology, Icahn School of Medicine at Mount Sinai, New York, New York, USA

†: Seyed Amir Ahmad Safavi-Naini and Mohammad Amin Khalafi contributed equally to this work.

*  Corresponding to: Hamid Asadzadeh Aghdaii (hamid.assadzadeh@gmail.com) and Ali Soroush (Ali.Soroush@mountsinai.org).

**Funding:** None
**Data Transparency Statement:** Code is available at:
https://github.com/aminkhalafi/CML-vs-LLM-on-Polyp-Detection
**Disclosure Statement:**
MAK: MAK has no conflict of interest related to this study.
SAASN: SAASN has no conflict of interest related to this study.
ND: ND has no conflict of interest related to this study.
DA: DA has no conflict of interest related to this study.
PKM: PKM has no conflict of interest related to this study.
AmSa: AmSa has no conflict of interest related to this study.
KK: KK has no conflict of interest related to this study.
NG: NG has no conflict of interest related to this study.
SS: SS has no conflict of interest related to this study.
SF: SF has no conflict of interest related to this study.
JSS: JSS has no conflict of interest related to this study.
NPT: NPT has no conflict of interest related to this study.

NH: NH as no conflict of interest related to this study.

GN: GN has no conflict of interest related to this study.

HAA: HAA has no conflict of interest related to this study.

AlSo: AlSo is on the advisory board and has equity in Virgo Surgical Solutions.

# Abstract

**Introduction:** Medical image analysis plays a pivotal role in clinical decision-making. Recent advancements in vision-language models (VLMs) offer promising capabilities for processing both visual and textual data simultaneously. This study provides a comprehensive performance assessment of VLMs against established convolutional neural networks (CNNs) and classic machine learning models (CMLs) for computer-aided detection (CADe) and computer-aided diagnosis (CADx) of colonoscopy polyp images.

**Method:** We analyzed 2,258 colonoscopy images with corresponding pathology reports from 428 patients. We preprocessed all images using standardized techniques (resizing, normalization, and augmentation) and implemented a rigorous comparative framework evaluating 11 distinct models: ResNet50, 4 CMLs (random forest, support vector machine, logistic regression, decision tree), two specialized contrastive vision language encoders (CLIP, BiomedCLIP), and three general-purpose VLMs ( GPT-4 Gemini-1.5-Pro, Claude-3-Opus). Our performance assessment focused on two clinical tasks: polyp detection (CADe) and classification (CADx).

**Result:** In polyp detection, ResNet50 achieved the best performance (F1: 91.35%, AUROC: 0.98), followed by BiomedCLIP (F1: 88.68%, AUROC: [AS1] ). GPT-4 demonstrated comparable effectiveness to traditional machine learning approaches (F1: 81.02%, AUROC: [AS2] ), outperforming other general-purpose VLMs.  For polyp classification, performance rankings remained consistent but with lower overall metrics. ResNet50 maintained the highest efficacy (weighted F1: 74.94%), while GPT-4 demonstrated moderate capability (weighted F1: 41.18%), significantly exceeding other VLMs (Claude-3-Opus weighted F1: 25.54%, Gemini 1.5 Pro weighted F1: 6.17%).

**Conclusion:** CNNs remain superior for both CADx and CADe tasks. However, VLMs like BioMedCLIP and GPT-4 may be useful for polyp detection tasks where training CNNs is not feasible.

# 1. Introduction

The interpretation of medical images represents a cornerstone of modern healthcare decision-making, with direct implications for patient diagnosis, treatment planning, and clinical outcomes. In gastroenterology, accurate detection and classification of colorectal polyps during colonoscopy is particularly crucial for effective cancer prevention [1]. Despite technological advancements, polyp identification remains inconsistent, even among experienced clinicians [2]. This clinical challenge reality has driven substantial interest in artificial intelligence (AI) systems that could augment clinical performance and standardize exam quality.

Recent advances in artificial intelligence have revolutionized automated medical image analysis. Traditional approaches leveraging convolutional neural networks (CNNs) such as VGG [3], Inception [4], ResNet [5], and DenseNet [6] have demonstrated considerable success in image classification tasks. However, creating clinically viable tools typically requires extensive labeled medical datasets, substantial computational resources, and specialized development for each clinical application. Recently, Contrastive Language-Image Pre-training (CLIP) has emerged as a robust approach capable of connecting textual descriptions directly to visual representations. This approach reduces or eliminates the need for large labeled datasets for each clinical task, which broadens the utility of AI image analysis [7].

The integration of vision and language capabilities through multimodal models introduces an additional paradigm shift in medical image analysis. By combining computer vision with natural language processing, vision-language models (VLMs) can process both visual and textual information, enabling more flexible interaction with medical images through natural language. Recent general-purpose VLMs such as GPT-4 [8], Claude-3-Opus [9], and Gemini-1.5-Pro [10] have demonstrated remarkable zero-shot capabilities across diverse domains without task-specific training. While preliminary studies have explored VLM performance in various medical domains (**Table 1**), comprehensive comparative analyses against established methodologies remain limited, particularly in gastroenterology. The question of whether these general-purpose models can approach or even match the performance of specialized architectures holds significant implications for clinical implementation, technological development, and healthcare accessibility.

**Table 1.** Overview of studies assessing the performance of vision language models in medical imaging.

| First Author, Year | VLM | Major | Modality | Performance |
|---|---|---|---|---|
| Pilia, 2024; and Hardin, 2024 [11] | GPT-4 | Dermatology | Image/ Scenario Prompt /Image + Scenario Prompt | GPT-4V accuracy: image-only: 54%/ text-only scenarios: 89%/ both image + scenario: 89% |
| Laohawetwanti, 2024 [12] | custom GPT-4 | Histopathology | Colorectal polyp photomicrographs | GPT-4 accuracy: 16% for non-specific changes / 36% for tubular adenomas Sensitivity: 74% for adenoma detection specificity: 36% for adenoma detection |
| Chen, 2023 [13] | GPT-4V | Internal medicine | COVID-19 lung X-ray | GPT4-V accuracy: ranged 72% to 85% based on different prompts. |
| Han, 2023 [14] | GPT-4 | General Medicine | Clinical cases from the JAMA Clinical Challenge and the NEJM Image Challenge | GPT-4V accuracy: 73.3% for JAMA and 88.7% for NEJM |
| Xu, 2024 [15] | GPT-4 | ophthalmology | various ocular imaging modalities | Examination Identification :95.6% Lesion Identification:25.6% Diagnosis Capacity:16.1% Decision Support:24% |
| Yang, 2023 [16] | GPT-4 | General Medicine | USMLE with Image | For questions with images: 86.2%, 73.1%, and 62.0% on USMLE, DRQCE, and AMBOSS. For questions with image, GPT-4 achieved an accuracy of 84.2%, 85.7%, 88.9% in Step1, Step2CK, and Step3 of USMLE questions |
| Jin, 2024 [17] | GPT-4 | General Medicine | Clinical cases from NEJM Image Challenges + scenario prompt | GPT-4 accuracy: 81.6%, which outperformed physicians and medical students. |

In this study, we provide a systematic head-to-head evaluation of eleven distinct computational approaches for colonoscopy polyp image analysis, including traditional machine learning models, specialized neural networks, and both general and domain-specific VLMs. Using a dataset of 2,258 colonoscopy images with pathological verification, we assess performance across two critical clinical tasks: polyp detection and histological classification. Our investigation provides additional insight into the comparative efficacy of these diverse methodologies, establishes performance benchmarks for future development, and offers evidence-based guidance for the potential integration of VLMs into gastroenterological practice.

## 2. Method

### 2.1. Ethical Consideration

This study received ethical approval from the Institutional Review Board at the Research Ethics Committees of the Research Institute for Gastroenterology & Liver Diseases at Shahid Beheshti University of Medical Sciences (Approval ID: IR.SBMU.RIGLD.REC.1401.043). In accordance with the principles outlined in the Helsinki Declaration, patient confidentiality and welfare was maintained throughout the study. All procedures involving patient data and images were conducted using standardized protocols to safeguard patient privacy, with measures in place to anonymize data and prevent identification. Explicit informed consent was obtained from all participants, affirming their voluntary participation in the study.

### 2.2. Experimental Framework

This investigation followed a retrospective, comparative methodological design to evaluate multiple artificial intelligence approaches for colonoscopy image analysis. We adhered to "Consolidated reporting guidelines for prognostic and diagnostic machine learning modeling studies" [11] and TRIPOD-AI [12] for model development and results reporting, ensuring methodological transparency and reproducibility. We structured our investigation as a three-phase experimental program designed to systematically evaluate model performance:

1. **Parameter Optimization Phase**: We systematically identified optimal hyperparameters for each model architecture through comprehensive grid search methodologies, establishing optimized configurations for subsequent performance evaluation.
2. **Detection Evaluation Phase**: We conducted comparative assessments of model performance in identifying polyp presence (CADe functionality), utilizing standardized metrics, including F1 scores and area under the receiver operating characteristic curve (AUROC).
3. **Classification Analysis Phase**: We performed systematic evaluation of model efficacy in correctly classifying polyp pathology types (CADx functionality) across six distinct histological categories, employing weighted evaluation metrics to account for class distribution.

This structured approach enabled comprehensive, controlled comparison across diverse computational methodologies while maintaining consistent evaluation standards.

## 2.3. Dataset - Characteristics

### 2.3.1. Patient Population and Data Collection

We examined colonoscopy data collected between December 2022 and April 2023 at Taleghani Hospital's gastroenterology clinic and Behbood clinic. The study population comprised 428 patients (mean age: 53±14 years; 48.6% male) who underwent colonoscopy for primary colorectal cancer screening, post-polypectomy surveillance, evaluation following positive fecal immunochemical tests, or investigation of gastrointestinal symptoms.

All procedures were performed by gastroenterologists with extensive experience (>2,000 screening colonoscopies conducted). The endoscopists assessed bowel preparation quality using the validated Boston Bowel Preparation Scale and confirmed cecal intubation through identification of the ileocecal valve and appendix orifice.

### 2.3.2. Image Collection and Histopathological Assessment

We compiled a comprehensive image dataset consisting of 1,129 colon polyp images and 1,129 randomly selected normal colon images (from an original pool of 6,046) to address class imbalance. The initial classification was derived from procedure pathology reports, followed by an expert review of stored images by an experienced gastroenterologist (PMK) who assigned final labels.

Tissue samples underwent standardized histopathological processing, including formalin fixation, paraffin embedding, sectioning (4-5 microns thick), and hematoxylin-eosin staining. Pathological classification followed established criteria [13], with an assessment of cellular atypia, glandular architecture, and dysplasia degree. The dataset distribution across pathological categories is presented in **Table 2**.

## 2.4. Image Preprocessing and Data Augmentation

We implemented a comprehensive preprocessing pipeline to optimize image quality and enhance model training. All images underwent uniform resizing to 300×300 pixels, followed by normalization to standardize pixel value distribution. To enhance model robustness and generalizability, we applied a systematic augmentation protocol incorporating horizontal and vertical mirroring to diversify polyp orientation representation, brightness variations to simulate diverse lighting conditions, Gaussian blur application to replicate optical aberrations, additive Gaussian noise to build resilience against image artifacts, and linear contrast adjustments to enhance structural differentiation. This augmentation strategy expanded the effective training dataset while promoting model generalizability across varying image acquisition conditions.

## 2.5. Model Development and Configuration

### 2.5.1 Classical Machine Learning Approaches

We implemented five distinct classical machine learning algorithms, each optimized through systematic hyperparameter tuning. For the Decision Tree Classifier, we employed a comprehensive grid search across multiple parameters, including criterion ('gini', 'entropy'),

max_depth (None, 10, 20, 30), min_samples_split (2, 5, 10), and min_samples_leaf (1, 2, 4). The optimal configuration identified was criterion='entropy', max_depth=20, min_samples_leaf=2, and min_samples_split=2. For the Random Forest Classifier, our hyperparameter optimization encompassed n_estimators (50, 100, 200), max_depth (None, 10, 20, 30), min_samples_split (2, 5, 10), and min_samples_leaf (1, 2, 4). The optimal configuration was determined to be n_estimators=200, min_samples_leaf=1, min_samples_split=10, and random_state=42. With the Support Vector Machine (SVM), we systematically evaluated parameter combinations including C (0.1, 1, 10), kernel ('linear', 'rbf', 'poly'), and gamma ('scale', 'auto'). The optimal configuration identified was kernel='rbf', C=10, gamma='scale', probability=True, and random_state=42. For Logistic Regression, our grid search evaluated C values (0.1, 1, 10) and solver options ('newton-cg', 'lbfgs', 'liblinear', 'sag', 'saga'). The optimal configuration was determined to be C=0.1, max_iter=100, solver='sag', and random_state=42. The Gaussian Naive Bayes algorithm was implemented with default parameters as it does not feature adjustable hyperparameters.

### 2.5.2. Convolutional Neural Network: Resnet 50

We implemented ResNet50 based on its demonstrated superior performance in medical image classification tasks [5]. To optimize performance, we conducted systematic hyperparameter tuning via GridSearchCV, evaluating learning_rate (0.01, 0.1, 1), epochs (5, 10, 15), and batch_size (32, 64). The grid search involved dividing the dataset into training, validation, and testing subsets, training the model on various hyperparameter combinations, and using cross-validation to evaluate performance and prevent overfitting. The optimal configuration was determined to be learning_rate=0.01, epochs=15, and batch_size=32.

### 2.5.3. Contrastive Multimodal Encoders

We evaluated two specialized contrastive learning models for our analysis. CLIP (Contrastive Language-Image Pretraining) represents a general-purpose multimodal model that associates images with corresponding textual descriptions through dual visual and textual encoders trained on 400 million image-text pairs [7]. We implemented the ViT-B/32 variant for zero-shot evaluation in our experimental framework. Additionally, we assessed BiomedCLIP, a domain-specialized adaptation of CLIP that underwent pretraining on PMC-15M—a dataset comprising 15 million biomedical figure-caption pairs from PubMed Central publications [14]. This biomedical specialization potentially enhances performance for medical imaging applications, making it particularly relevant for our colonoscopy image analysis.

### 2.5.4. General-Purpose Vision Language Models

We evaluated three state-of-the-art VLMs as part of our comprehensive assessment. GPT-4 represents an enhanced iteration of OpenAI's GPT-4 model that integrates advanced visual processing capabilities, enabling interpretation of and response to image inputs [15]. We also included Claude-3-Opus, developed by Anthropic, which builds upon their Claude architecture with enhanced visual question answering capabilities [9]. The third model in our evaluation was Gemini-1.5-Pro, Google's multimodal foundation model designed for versatile tasks including visual comprehension, classification, and content generation across modalities [10]. These general-purpose models were evaluated without domain-specific fine-tuning to assess their zero-shot capabilities in medical image analysis.

We utilized our experiments' web interfaces of GPT-4 (*gpt-4-1106-vision-preview;* Accessed: May 2024 via API), Claude-3-Opus (*claude-3-opus-20240229;* Accessed: May 2024 via API ), and Gemini-1.5-Pro (*gemini-1.5-pro-001;* Accessed: June 2024 via Google interface), accessed during April-June 2024. Approximately 15% of our test dataset was allocated for Experiment 0, while the remaining 85% was used for Experiments 1 and 2. All experiments were conducted with standardized parameters (temperature = 1.0, maximum tokens = 512, tool calls disabled, random seed = 123) to ensure consistent evaluation conditions. We renamed all image file names to avoid any data leakage from the image metadata.

In Experiment 0, we used the following raw prompt in a chat: "*What is this image?*" accompanied by the image. Subsequently, in the same chat, we asked: "*What is the pathology class of the polyp? Give me only one answer.*" In a separate chat for Experiment 0, we posed this best prompt:

> "*As an esteemed gastroenterologist specializing in colonoscopy evaluation, your expertise is crucial in meticulously assessing a provided colonoscopy image. Your task is* *to discern and characterize any irregularities present across the colonic mucosa, paying close attention to morphology, color variations, and vascularity patterns. Drawing upon your wealth of experience, construct a comprehensive list of potential diagnoses, including but not limited to inflammatory bowel disease, colorectal polyps, diverticulosis, and colorectal cancer. Your discerning analysis and diagnostic acumen will guide subsequent clinical decisions, emphasizing the importance of accurate interpretation and effective communication in delivering optimal patient care."*

This was followed by the image. Then, in the same chat, we used the prompt:

*"Analyze the provided image and select one of the following options that accurately describes the patient's diagnosis:*

1. *normal*
2. *adenocarcinoma*
3. *adenomatous-tubular polyp*
4. *adenomatous-tubulovillous polyp*
5. *adenomatous-villous polyp*
6. *hyperplastic polyp*
7. *inflammatory polyp."*

## 2.6. Performance Evaluation

We developed an approach that converts unstructured text into structured classifications using GPT-4 to facilitate the semi-automated evaluation of textual outputs. The model was configured with a temperature setting of 0 and enabled to generate structured JSON outputs. It was designed to categorize responses into predefined labels, including: (1) "Human evaluation needed: I am unsure," (2) "Human evaluation needed: More than one diagnosis is selected, or no option is selected," (3) "The unstructured answer selected: No polyp is detected in the image," (4) "The unstructured answer selected: A polyp is detected in the image," and (5) "The unstructured answer selected: The polyp type is classified as {polyp_type in polyp_types}." This approach enables automated classification while flagging ambiguous or uncertain cases for human review, ensuring accuracy in the structured evaluation process. In a random sample of 50 response-extraction pairs, GPT-4 correctly labeled all 43 extractions while labeled 7 for human evaluation.

## 2.7. Statistical Analysis

We performed comprehensive statistical analysis using Python (version 3.11.5), employing standardized machine learning evaluation methodologies. We implemented the one-vs-all strategy for multiclass classification scenarios to enable binary performance metrics for each class. We selected the F1 score as our primary evaluation metric due to its balanced consideration of both precision and recall, making it particularly suitable for our dataset where class imbalance was present, especially in the polyp classification tasks where some pathology types had limited representation.

Performance was evaluated using multiple complementary metrics: F1 scores to balance precision and recall considerations; AUROC to assess discriminative capability; confusion matrices to visualize classification patterns and error types; and weighted metrics to account for class imbalance in overall performance assessment. For weighted F1 scores in polyp classification tasks, we calculated values based on the proportion of each polyp type in the test dataset, ensuring that performance metrics appropriately reflected the distribution of classes in clinical settings.

## 2.8. TiLense: Importance of Tiles for VLM's Zero-Shot Polyp Detection

This proposed approach seeks to identify and visualize key image tiles in vision-language tasks by assessing the significance of each tile through frequent responses across multiple prediction attempts. In contrast to complex methods, it focuses on a single, dominant answer instead of the original model probability. The procedure involves pinpointing the primary answer, evaluating tile significance, and then generating a heatmap to showcase these important areas. This model-agnostic unsupervised technique elucidates essential regions in VLM classification by juxtaposing tile-based results with a singular base answer after N iterations. By highlighting areas where significant variations occur, it uncovers which sections of an image most influence model predictions, which is beneficial for evaluation and improvement. We refer to this method as "TiLense" due to its capacity to highlight importance across image tiles for zero-shot prediction tasks.

We implemented this tile masking technique to showcase GPT-4's vision capabilities in zero-shot prediction tasks across four scenarios: the presence of a polyp, a polyp in a challenging background, a standard image, and a standard image in a complex background. A systematic sliding window approach masked specific regions of the images (see Figure 1). The original and masked images were evaluated by GPT-4 using a standardized prompt, with a temperature setting of 1, a maximum token limit of 300, and no specified seed value, with the process repeated five times to create response distributions. The base answer were established through majority voting. The output is represented as a heatmap, where each tile is colored according to its impact on altering the base answer. Since tiles can overlap, we scale each tile from 0 to 1, coloring them from white to red.

# 3. Results

## 3.1. Model Optimization

Our initial experimental phase focused on optimizing model configurations and prompt strategies for vision language models. We observed that domain-specific prompts consistently outperformed simple queries across all VLMs tested. GPT-4 demonstrated a 17.6% improvement in F1 score for polyp detection with optimized prompts (from 0.636 to 0.748), while Claude-3-Opus showed a more substantial 72.2% improvement (from 0.266 to 0.458). For polyp classification, the impact was even more pronounced, with GPT-4 achieving a 434.9% improvement in F1 score using the specialized prompt (from 0.126 to 0.548), and Claude-3-Opus showing a 31.2% improvement (from 0.112 to 0.147). **Table 3** provides a detailed comparison of performance improvements across prompting strategies, which formed the foundation for our subsequent analyses. **Supplementary Figures S1** and **S2** present the confusion matrices of answers for polyp detection and classification.

**Table 3.** Impact of prompt engineering on vision language model performance.

|  | GPT-4 | Claude-3-Opus | Gemini-1.5-Pro |
|---|---|---|---|
|  | F1 score (change) | F1 score (change) | F1 score (change) |
| **Simple prompt for polyp detection** | 0.636 (ref) | 0.266 (ref) | 0.715 (ref) |
| **Engineered prompt for polyp detection** | 0.748 (+17.6%) | 0.458 (+72.2%) | 0.731 (+2.2%) |
| **Simple prompt for polyp classification** | 0.126 (ref) | 0.112 (ref) | 0.0 (ref) |
| **Engineered prompt for polyp classification** | 0.548 (+434.9%) | 0.147 (+31.2%) | 0.437 (NA) |

## 3.2 Polyp Detection Performance (CADe)

We established a clear performance hierarchy in polyp detection capability across model architectures (**Figure 1**, **Table 4**). ResNet50 demonstrated superior performance (F1: 91.35%, AUROC: 0.98), followed by BiomedCLIP (F1: 88.68%). Traditional machine learning approaches and GPT-4 formed the next performance tier, with Random Forest and GPT-4 both achieving identical F1 scores of 81.02%, followed by SVM (F1: 77.92%, AUROC: 0.84) and Logistic Regression (F1: 72.80%, AUROC: 0.82). Decision Tree (F1: 68.10%, AUROC: 0.75) and CLIP (F1: 68.39%, AUROC: 0.76) showed moderate capability, while Claude-3-Opus demonstrated limited effectiveness (F1: 66.40%, AUROC: 0.71). Gemini 1.5 Pro and Gaussian Naive Bayes exhibited the lowest detection capability (F1: 19.37% and 10.22%, AUROC: 0.52 and 0.51, respectively). Confusion matrices for polyp detection across all models are presented in **Figure 1**, illustrating the distribution of true positives, true negatives, false positives, and false negatives for each approach. The AUROC values for polyp detection across all models are illustrated in **Figure 2**.

**Table 4.** Comparative Analysis of Machine Learning Models in Polyp Detection and Classification. Performance comparison of Classical Machine Learning (CML) models, ResNet-50, Vision Language Models (VLMs), and specialized VLMs for polyp detection and classification tasks. (a) F1 scores demonstrating model performance in polyp detection and classification across different architectures. The bolded values represent the highest for each task in the column. Abbreviations: CML, Classical Machine Learning; VLM, Vision Language Model; cVL, contrastive Vision-Language encoders; AC, Adenocarcinoma; TA, Tubular Adenoma; TVA, Tubulovillous Adenoma; VA, Villous Adenoma; HP, Hyperplastic Polyp; IP, Inflammatory Polyp.

| Model Family | Model | Polyp Detection | Polyp Classification | AC (N=79) | TA (N=771) | TVA (N=59) | VA (N=59) | HP (N=138) | IP (N=45) |
|---|---|---|---|---|---|---|---|---|---|
| | | F1 | Weighted F1 | F1 | F1 | F1 | F1 | F1 | F1 |
| CML | Decision tree | 0.681 | 0.4042 | 0.06 | 0.53 | 0.27 | 0.00 | 0.22 | 0.14 |
| CML | Random forest | 0.8102 | 0.4367 | 0.00 | 0.64 | 0.00 | 0.00 | 0.08 | 0.00 |
| CML | Support vector machine | 0.7792 | 0.5563 | 0.45 | 0.68 | 0.25 | 0.00 | 0.31 | 0.36 |
| CML | Logistic regression | 0.728 | 0.4032 | 0.10 | 0.56 | 0.00 | 0.00 | 0.07 | 0.20 |
| CML | Gaussian naive bayes | 0.1022 | 0.0764 | 0.08 | 0.09 | 0.00 | 0.00 | 0.07 | 0.00 |
| CNN | ResNet50 | 0.9135 | 0.7494 | **0.67** | **0.85** | **0.55** | **0.25** | **0.49** | **0.71** |
| VLM | GPT-4 | 0.8102 | 0.4118 | 0.30 | 0.58 | 0.00 | 0.00 | 0.00 | 0.00 |
| VLM | Claude-3-Opus | 0.664 | 0.2554 | 0.19 | 0.33 | 0.06 | 0.04 | 0.14 | 0.00 |
| VLM | Gemini-1.5-Pro | 0.1937 | 0.0617 | 0.00 | 0.09 | 0.00 | 0.00 | 0.00 | 0.00 |
| cVL | CLIP | 0.6839 | 0.0169 | 0.19 | 0.00 | 0.00 | 0.00 | 0.00 | 0.04 |
| cVL | BiomedCLIP | 0.8868 | 0.2774 | 0.56 | 0.29 | 0.17 | 0.00 | 0.21 | 0.04 |

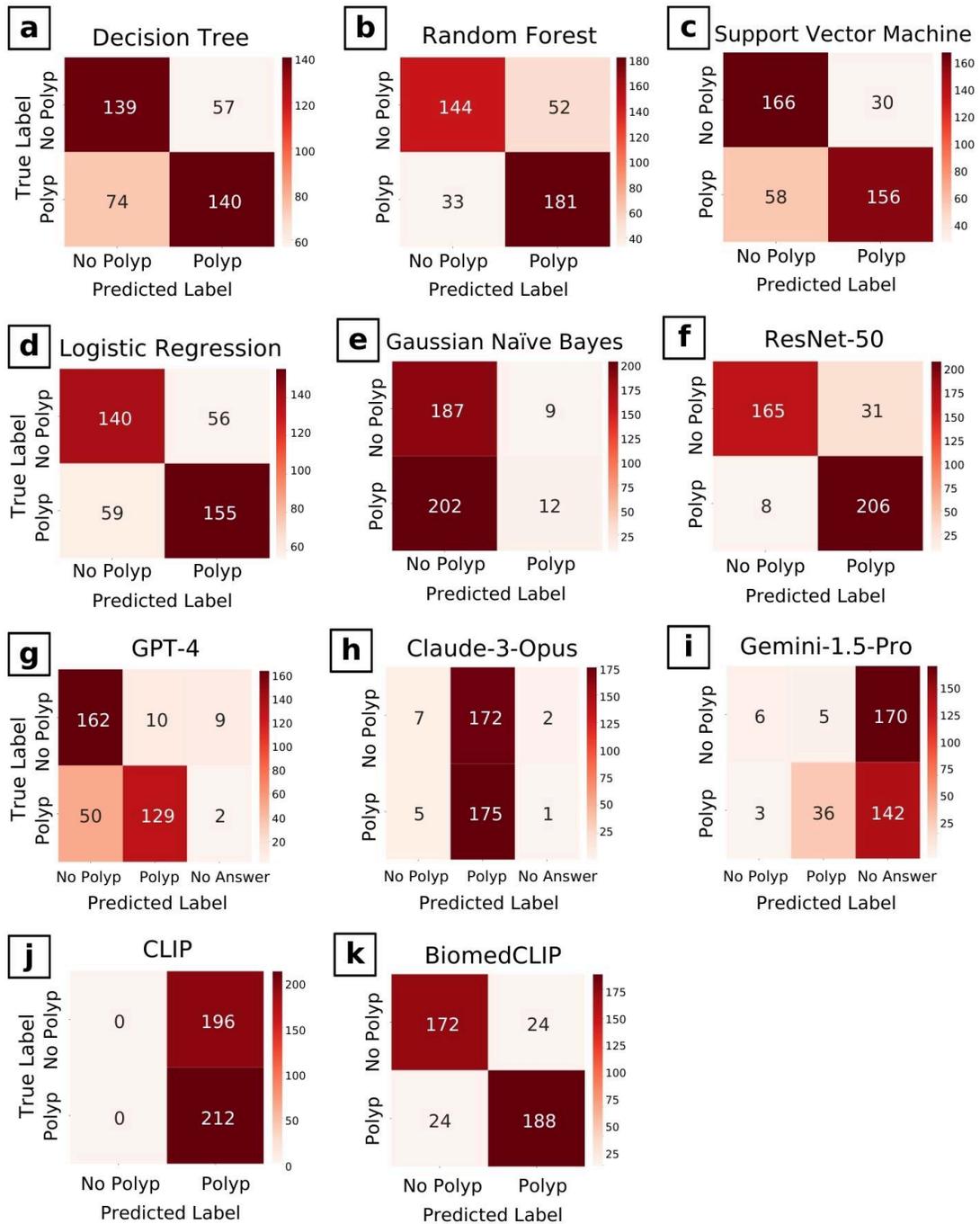

**Figure 1.** Confusion matrices depicting polyp detection performance across various models: classical machine learning algorithms—Decision Tree (**a**), Random Forest (**b**), Support Vector Machine (**c**), Logistic Regression (**d**), Gaussian Naïve Bayes (**e**); convolutional neural network—ResNet-50 (**f**); vision-language models—GPT-4 (**g**), Claude-3-Opus (**h**), Gemini-1.5-Pro (**i**); and contrastive vision-language encoders—CLIP (**j**), BiomedCLIP (**k**). Each matrix illustrates model predictions relative to ground-truth labels.

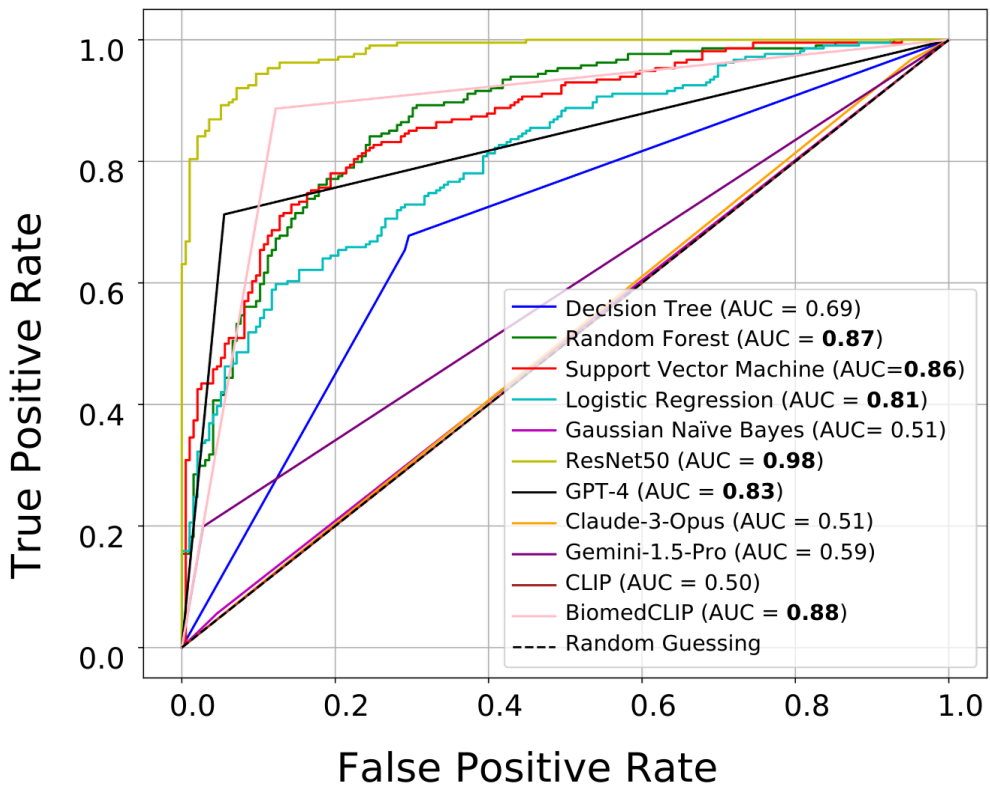

**Figure 2.** Receiver operating characteristic curves for polyp detection, with the corresponding area under the curve (AUC) values. AUC values greater than 0.8 are shown in bold.

**Figure 3** presents the heatmap of tile importance resulting from the integration of the TiLense approach (**Figure 3.a**) into GPT-4, analyzed across both standard images (**Figures 3.b** and **3.c**) and more challenging images (**Figures 3.d and 3.e**). The TiLense approach effectively elucidates zero-shot VLM insights for positive and localized findings (**Figure 3.b**), though it encounters difficulties when explaining standard images (**Figure 3.b**). GPT-4 is susceptible to being misled in challenging cases, as evidenced by the hard-to-see polyp in **Figure 3.d** and the poorly prepared normal image in **Figure 3.e**.

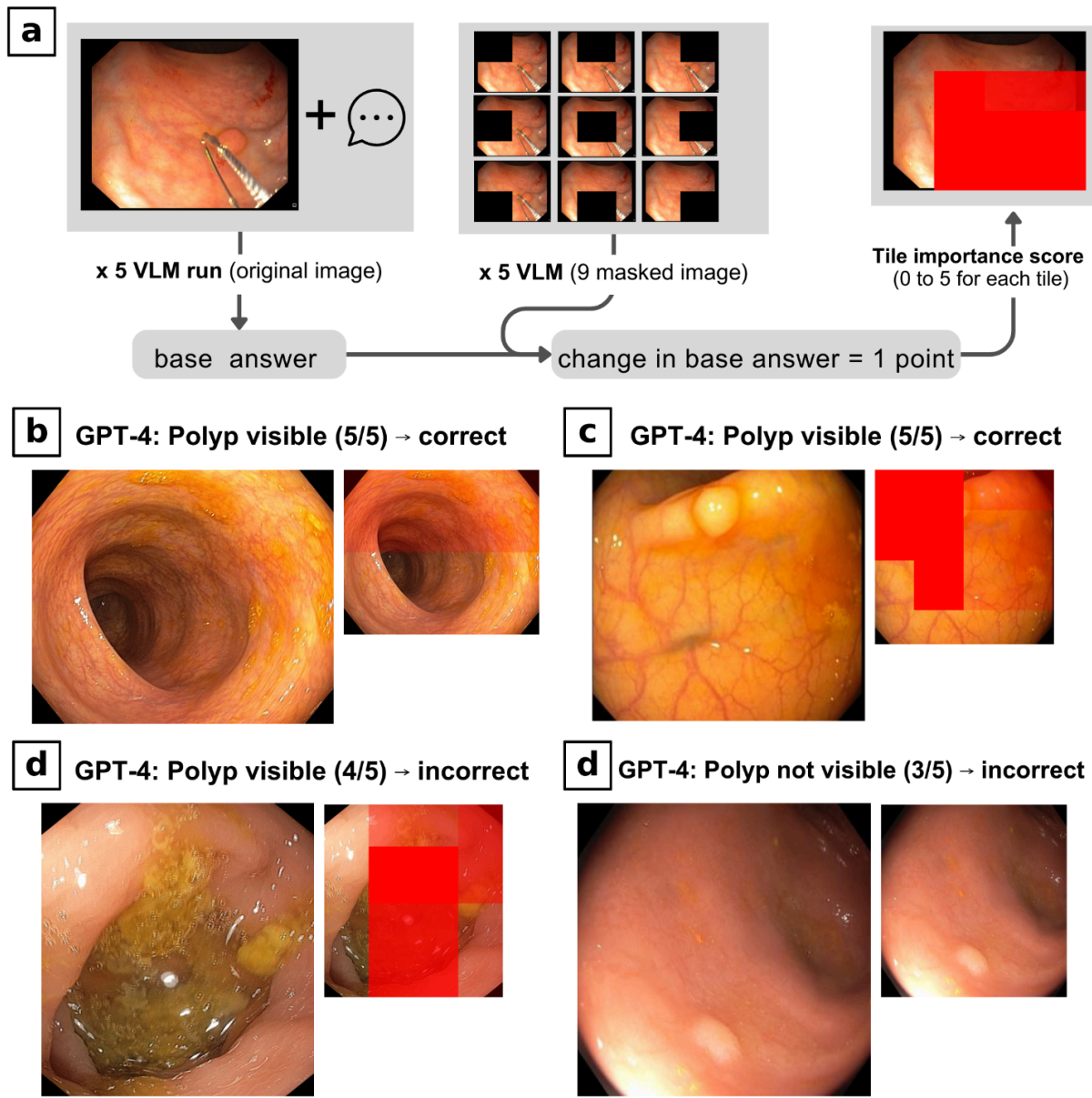

**Figure 3.** Evaluation of GPT-4 for polyp detection using TiLense, focusing on tile-level importance. The method includes five runs with vision-language models (VLMs) on original and masked images, using 9 masked tiles per image. Each tile receives an importance score from 0 to 5, indicated by a color gradient from white to red, where red denotes a tile whose removal alters the base answer significantly. A reference answer for each image is established, and deviations are scored as 1 point. Panels (a–e) show tile-level predictions across image conditions: standard image without polyp (b), standard image with polyp (c), challenging image without polyp (d), and challenging image with polyp (e).

### 3.3 Polyp Classification Performance (CADx)

The performance of classification varied significantly based on model architecture and polyp types (**Figure 4**). **Table 4** presents the weighted F1 scores for polyp classification, while **Supplementary Table S1** details the performance categorized by polyp type. ResNet50 consistently achieved the highest performance across all histological categories. The overall weighted classification performance demonstrated clear stratification: ResNet50 (74.94%), SVM (55.63%), Random Forest (43.67%), GPT-4 (41.18%), BiomedCLIP (27.74%), Claude-3-Opus (25.54%), Gemini 1.5 Pro (6.17%), and CLIP (1.69%).

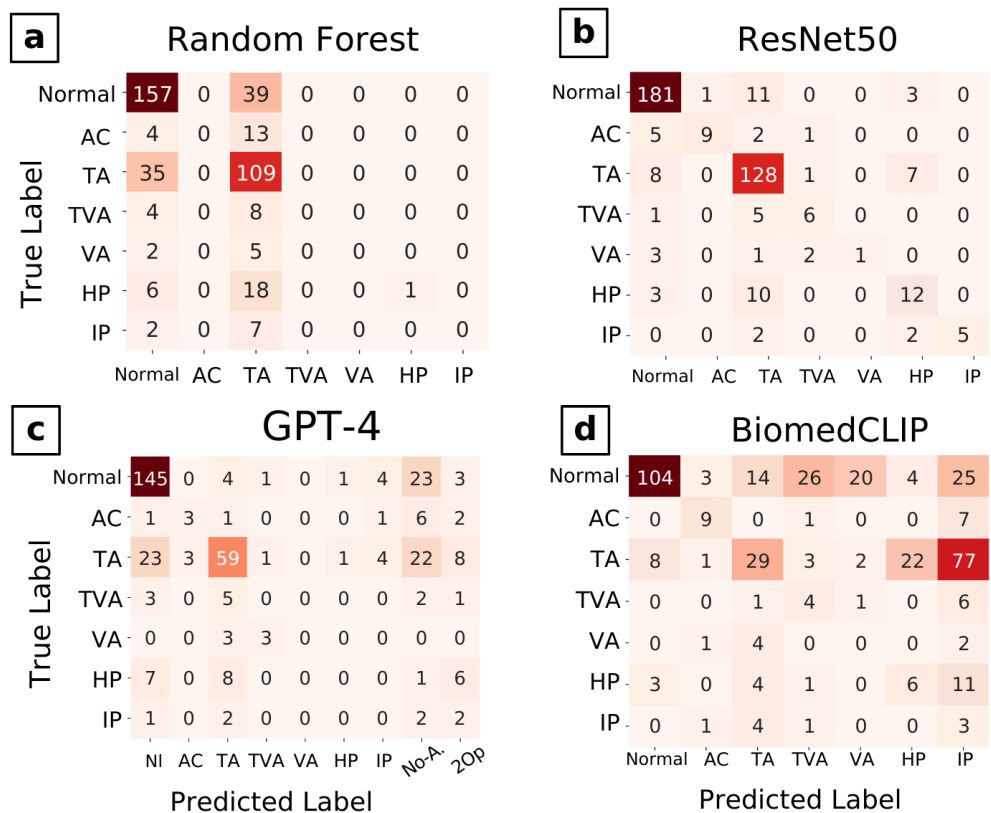

**Figure 4.** Confusion matrices of polyp classification are provided for the top-performing classical machine learning model (a: Random Forest), convolutional neural network (b: ResNet-50), highest-performing vision-language mode (c: GPT-4a), and the contrastive vision-language encoder fine-tuned on external general medical imaging data (d: BiomedCLIP. Abbreviations: AC, Adenocarcinoma; TA, Tubular Adenoma; TVA, Tubulovillous Adenoma; VA, Villous Adenoma; HP, Hyperplastic Polyp; IP, Inflammatory Polyp; No-A: No answer provided; 2OP: two options (polyp type) were selected.

Tubular adenoma images (650 training, 121 test) showed the highest overall classification performance across models. ResNet50 achieved an F1 score of 84.77% (AUROC: 0.94), with SVM (F1: 67.52%, AUROC: 0.75) and Random Forest (F1: 63.56%, AUROC: 0.71) showing strong capability. GPT-4 demonstrated moderate effectiveness (F1: 58.13%, AUROC: 0.70), significantly outperforming other VLMs, including Claude-3-Opus (F1: 32.72%, AUROC: 0.64).

For adenocarcinoma images (66 training, 13 test), ResNet50 achieved an F1 score of 66.67% (AUROC: 0.99), with BiomedCLIP showing moderate capability (F1: 56.25%, AUROC: 0.76) and SVM performing adequately (F1: 45.45%, AUROC: 0.65). GPT-4 (F1: 30.00%, AUROC: 0.60) outperformed Claude-3-Opus (F1: 18.92%, AUROC: 0.67) and Gemini-1.5-Pro (F1: 0%, AUROC: 0.50).

Hyperplastic polyp images (116 training, 22 test) presented a more challenging classification task. ResNet50 demonstrated an F1 score of 48.98% (AUROC: 0.95), with SVM showing limited effectiveness (F1: 31.25%, AUROC: 0.61) and Decision Tree and BiomedCLIP performing marginally (F1: 21.62% and 21.05%, AUROC: 0.57 and 0.59, respectively).

The most challenging classifications were observed for tubulovillous adenoma (48 training, 11 test), villous adenoma (30 training, 6 test), and inflammatory polyp images (38 training, 7 test). For tubulovillous adenoma, ResNet50 achieved an F1 score of 54.55% (AUROC: 0.96), while most other models performed at or near random chance. For villous adenoma, even ResNet50 showed limited capability (F1: 25.00%, AUROC: 0.88), with all other models demonstrating minimal effectiveness. For inflammatory polyp images, ResNet50 demonstrated relatively strong performance (F1: 71.43%, AUROC: 0.95), while SVM achieved moderate success (F1: 36.36%, AUROC: 0.61) and other approaches showed limited capability.

**Figure 2** displays confusion matrices for polyp classification utilizing Random Forest (CML's top performer), ResNet50, GPT-4 (the leading VLM), and BiomedCLIP. These matrices indicate persistent misclassification tendencies, especially among similar adenoma subtypes, which pose the most significant challenge for all methods assessed. Additionally, the ROC curve and confusion matrices for all models can be found in **Supplementary Figures S3** and **S4**.

# 4. Discussion

This study provides a comprehensive comparison of VLMs and traditional AI approaches for polyp detection and classification in colonoscopy images. Our findings contribute to understanding the capabilities and limitations of these computational paradigms for specific gastroenterology tasks.

Our systemic evaluation using standardized metrics established a performance hierarchy across computational paradigms tested The specialized CNN architecture (ResNet50) demonstrated superior performance for both polyp detection (F1: 91.35%) and classification (weighted F1: 74.94%). Among VLMS, GPT-4 emerged as the strongest VLM performer, achieving polyp detection (F1: 81.02%) comparable to traditional machine learning approaches. This direct comparison addresses an important methodological question about the relative capabilities of these different approaches under controlled conditions.

Our systematic evaluation established a clear performance hierarchy across computational paradigms. The specialized CNN architecture (ResNet50) consistently demonstrated superior performance for both polyp detection (F1: 91.35%) and classification (weighted F1: 74.94%). GPT-4 emerged as the strongest VLM performer, achieving detection capabilities (F1: 81.02%) comparable to traditional machine learning approaches. This finding quantifies for the first time the relative performance relationship between these computational paradigms specifically for coloscopy applications, addressing a significant knowledge gap in the medical AI literature.

The performance variations among VLMs were substantial and exceeded those typically observed in general image tasks, suggesting significant gaps in medical image processing capabilities. Though GPT-4 demonstrated moderate effectiveness for polyp detection (F1: 81.02%), Claude-3-Opus (F1: 66.40%) and Gemini-1.5-Pro (F1: 19.37%) performed considerably worse on our dataset. These findings align with emerging research comparing VLM performance across various medical imaging tasks from other clinical domains [16–21]. The significant performance disparities observed here may reflect variations in training data composition, model architecture and fine-tuning, or evaluation methodology and highlight the importance of systematic evaluation before clinical implementation of VLMs.

The performance gap between the domain-adapted BiomedCLIP (F1: 88.68% for detection) and the general-purpose CLIP (F1: 68.39%) reiterates the benefit of domain-specific pretraining in medical image analysis. In this case, BiomedCLIP was trained on additional 15 million biomedical figure-caption pairs, enabling the model to learn domain-specific visual features and their associations with medical terminology. Recent studies have demonstrated the effectiveness of this transfer learning approach across diverse medical imaging tasks[22].

Our analysis of prompt engineering effectiveness revealed striking performance sensitivity to prompt design, with GPT-4 showing a 17.6% improvement in detection and an 434.9% improvement in classification performance when using optimized prompts compared to simple queries. This exceeds previously reported prompt-dependent performance variations and reinforces that effective prompt engineering is critical for clinical VLM implementation[23,24]. This finding has significant implications for clinical deployment, indicating that consistent and well-designed prompts are essential to effective medical VLM applications. Newer VLM models may be less prone to prompt-dependent performance gaps [25–27].

A key finding was the differential performance across histological categories. VLM performance declining substantially for less common polyp types in our dataset. GPT-4's F1 score dropped from 58.13% for common tubular adenomas to 0% for rarer polyps like villous and tubulovillous adenomas. Similarly, Claude-3-Opus achieved F1 scores of 32.72% for tubular adenomas and 3.64% for villous adenomas. This pattern was more pronounced in VLMs than in traditional machine learning approaches, suggesting potential challenges with generalization to uncommon presentations despite these models' large-scale pretraining. This performance variation has important implications for potential clinical applications, particularly for tasks requiring reliable identification of rare but clinically significant findings.

Several methodological limitations should be considered when interpreting our findings. First, despite efforts to balance our dataset, the natural prevalence disparities across polyp types inevitably influenced classification performance. Second, our evaluation using still images rather than video sequences does not fully capture the real-world clinical image tasks performed during colonoscopies. Third, while we demonstrated substantial improvement with our domain-specific prompts, a more comprehensive exploration of prompting strategies could potentially yield further performance gains. Fourth, our study focused on performance metrics rather than interpretability or explainability, which are important considerations for clinical implementation.

From a clinical implementation perspective, our findings highlight the need for tailored deployment strategies based on specific clinical requirements and resource constraints. For high-stakes diagnostic applications that require precision, CNNs like ResNet50 remain the preferred choice when sufficient training data and technical resources are available, as we previously demonstrated with tabular data for zero-shot predictions[28]. However, VLMs may present viable alternatives in certain contexts, especially for polyp detection tasks where a traditional machine learning model may be impractical due to the absence of labeled data. The significant performance gap in classification tasks suggests that VLMs should not be considered for diagnostic classification without further advancements in the underlying technology

Future research could address several limitations identified in this study. First, investigating the effects of pretraining VLMs with comprehensive domain-specific data might help bridge the performance gaps with CNNs. Second, VLM performance on rare polyp types may improve with improved training data composition, synthetic data augmentation, or few-shot learning adaptations. Third, extending evaluation to video sequences could better approximate real-world clinical colonoscopy. Finally, assessing model interpretability would contribute to understanding how these systems arrive at their predictions, which is particularly important for clinical trust and implementation.

## 5. Conclusion

This comprehensive evaluation establishes a performance hierarchy among AI approaches for colonoscopy polyp detection and classification, with specialized CNNs leading in accuracy, followed by domain-adapted VLMs, and then general-purpose VLMs. These findings reinforce that specialized architectures like ResNet50 should remain the standard for high-stakes diagnostic applications. However, the limited zero-shot capabilities of properly engineered VLMs may be useful in scenarios where accuracy is less critical and creating a tailored CNN is infeasible. The dramatic performance improvements achieved through domain-specific pretraining and prompt optimization highlight practical enhancement pathways that could accelerate clinical adoption of VLMs. Looking forward, addressing VLMs' vulnerability to class imbalance and extending evaluations to video colonoscopy are critical. As these technologies continue to evolve, they hold promise to enhance standardization, reduce missed lesions, and ultimately improve colonoscopy quality across diverse healthcare settings.

## Conflict of Interests Declaration

AlSo serves on the advisory board and holds equity in Virgo Surgical Solutions. The other authors declare no conflicts of interest.

## Acknowledgments

The authors utilized ChatGPT, Claude, and Grammarly to assist with manuscript editing. All authors have reviewed the manuscript and take full responsibility for its content.

## Authors' Contribution

## Availability of Data

The datasets created and analyzed in this study cannot be accessed publicly due to IRB requirements; however, they can be obtained from the corresponding author (HAA) and SAASN (sdamirsa@gmail.com) upon a reasonable request after providing the IRB code. The code for the generation and evaluation of responses is publicly available at: https://github.com/aminkhalafi/CML-vs-LLM-on-Polyp-Detection

This is a supplementary file to "**Vision Language Models versus Machine Learning Models Performance on Polyp Detection and Classification in Colonoscopy Images**" by Mohammad Amin Khalafi, Seyed Amir Ahmad Safavi-Naini, Ameneh Salehi, Nariman Naderi, Dorsa Alijanzadeh, Pardis Ketabi Moghadam, Kaveh Kavosi, Negar Golestani, Shabnam Shahrokh, Soltanali Fallah, Jamil S Samaan, Nicholas P. Tatonetti, Nicholas Hoerter, Girish Nadkarni, Hamid Asadzadeh Aghdaei*, Ali Soroush*

Corresponding to Hamid Asadzadeh Aghdaii (hamid.assadzadeh@gmail.com) and Ali Soroush (Ali.Soroush@mountsinai.org).

## List of Supplementary

**Supplementary Table S1.** The accuracy and binary (one-vs-all) confusion matrix of polyp detection and polyp classification, stratified by type of polyp. Abbreviations: CML, classic machine learning; VLM: vision language model; cVL: contrastive vision language encoder; SVM: Support Vector Machine.

**Supplementary Figure S1.** The confusion matrix of polyp detection elucidates the results obtained from the simple prompt (left column: GPT-4 (a), Claude-3-Opus (c), Gemini-1.5-Pro (d)) alongside the outcomes produced from the engineered prompt (right column: GPT-4 (b), Claude-3-Opus (d), Gemini-1.5-Pro (f)).

**Supplementary Figure S2.** The confusion matrix of polyp classification elucidates the results obtained from the simple prompt (left column: GPT-4 (a), Claude-3-Opus (c), Gemini-1.5-Pro (d)) alongside the outcomes produced from the engineered prompt (right column: GPT-4 (b), Claude-3-Opus (d), Gemini-1.5-Pro (f)). Abbreviations: AC, Adenocarcinoma; TA, Tubular Adenoma; TVA, Tubulovillous Adenoma; VA, Villous Adenoma; HP, Hyperplastic Polyp; IP, Inflammatory Polyp; No-A: No Answer provided; 2OP: Two options (polyp type) is selected.

**Supplementary Figure S3.** The obtained receiver operating characteristic curve and the area under the curve (AUC) for the classification of polyp types, including adenocarcinoma (a), tubular adenoma (b), tubulovillous adenoma (c), villous adenoma (d), hyperplastic polyp (e), and inflammatory polyp (f).

**Supplementary Figure S4.** Confusion matrices depicting polyp classification performance across various models: classical machine learning algorithms—Decision Tree (a), Random Forest (b), Support Vector Machine (c), Logistic Regression (d), Gaussian Naive Bayes (e); convolutional neural network—ResNet-50 (f); vision-language models—GPT-4 (g), Claude-3-Opus (h), Gemini-1.5-Pro (i); and contrastive vision-language encoders—CLIP (j), BiomedCLIP (k). Each matrix illustrates model predictions relative to ground-truth labels.

**Supplementary Table S1.** The accuracy and binary (one-vs-all) confusion matrix of polyp detection and polyp classification, stratified by type of polyp. Abbreviations: CML, classic machine learning; VLM: vision language model; cVL: contrastive vision language encoder; SVM: Support Vector Machine.

| | | Polyp Detection | Adeno-carcinoma | Tubular adenoma | Tubulovillous adenoma | Villous adenoma | Hyperplastic polyp | Inflammatory polyp | Weighted F1 score |
|---|---|---|---|---|---|---|---|---|---|
| CML: Decision Tree | F1 | 0.681 | 0.057 | 0.525 | 0.267 | 0 | 0.216 | 0.143 | 0.404 |
| | TP | 140 | 1 | 73 | 4 | 0 | 4 | 1 | |
| | FP | 57 | 17 | 61 | 14 | 4 | 8 | 4 | |
| | TN | 139 | 376 | 205 | 384 | 399 | 377 | 397 | |
| | FN | 74 | 16 | 71 | 8 | 7 | 21 | 8 | |
| CML: Random Forest | F1 | 0.81 | 0 | 0.636 | 0 | 0 | 0.077 | 0 | 0.437 |
| | TP | 181 | 0 | 109 | 0 | 0 | 1 | 0 | |
| | FP | 52 | 0 | 90 | 0 | 0 | 0 | 0 | |
| | TN | 144 | 393 | 176 | 398 | 403 | 385 | 401 | |
| | FN | 33 | 17 | 35 | 12 | 7 | 24 | 9 | |
| CML: SVM | F1 | 0.779 | 0.455 | 0.675 | 0.25 | 0 | 0.313 | 0.364 | 0.556 |
| | TP | 156 | 5 | 106 | 2 | 0 | 5 | 2 | |
| | FP | 30 | 0 | 64 | 2 | 0 | 2 | 0 | |
| | TN | 166 | 393 | 202 | 396 | 403 | 383 | 401 | |
| | FN | 58 | 12 | 38 | 10 | 7 | 20 | 7 | |
| CML: Logistic Regression | F1 | 0.728 | 0.1 | 0.563 | 0 | 0 | 0.067 | 0.2 | 0.403 |
| | TP | 155 | 1 | 89 | 0 | 0 | 1 | 1 | |
| | FP | 56 | 2 | 83 | 3 | 1 | 4 | 0 | |
| | TN | 140 | 391 | 183 | 395 | 402 | 381 | 401 | |
| | FN | 59 | 16 | 55 | 12 | 7 | 24 | 8 | |
| CML: GNB | F1 | 0.102 | 0.077 | 0.092 | 0 | 0 | 0.071 | 0 | 0.076 |

| | | Polyp Detection | Adeno-carcinoma | Tubular adenoma | Tubulovillous adenoma | Villous adenoma | Hyperplastic polyp | Inflammatory polyp | Weighted F1 score |
|---|---|---|---|---|---|---|---|---|---|
| | TP | 12 | 1 | 7 | 0 | 0 | 1 | 0 | |
| | FP | 9 | 8 | 1 | 1 | 3 | 2 | 1 | |
| | TN | 187 | 385 | 265 | 397 | 400 | 383 | 400 | |
| | FN | 202 | 16 | 137 | 12 | 7 | 24 | 9 | |
| CNN: ResNet50 | F1 | 0.914 | 0.667 | 0.848 | 0.546 | 0.25 | 0.49 | 0.714 | 74.94 |
| | TP | 206 | 9 | 128 | 6 | 1 | 12 | 5 | |
| | FP | 31 | 1 | 30 | 4 | 0 | 12 | 0 | |
| | TN | 165 | 392 | 236 | 394 | 403 | 373 | 401 | |
| | FN | 8 | 8 | 16 | 6 | 6 | 13 | 4 | |
| VLM: GPT4 | F1 | 0.81 | 0.3 | 0.581 | 0 | 0 | 0 | 0 | 0.411 |
| | TP | 129 | 3 | 59 | 0 | 0 | 0 | 0 | |
| | FP | 10 | 3 | 23 | 5 | 0 | 2 | 9 | |
| | TN | 162 | 345 | 218 | 346 | 356 | 338 | 346 | |
| | FN | 50 | 11 | 62 | 11 | 6 | 22 | 7 | |
| VLM: Claude–Opus | F1 | 0.664 | 0.189 | 0.327 | 0.057 | 0.036 | 0.143 | 0 | 0.255 |
| | TP | 175 | 7 | 36 | 1 | 1 | 9 | 0 | |
| | FP | 172 | 53 | 63 | 23 | 48 | 95 | 14 | |
| | TN | 7 | 295 | 178 | 328 | 308 | 245 | 341 | |
| | FN | 5 | 7 | 85 | 10 | 5 | 13 | 7 | |
| VLM: Gemini-1.5-Pro | F1 | 0.194 | 0 | 0.092 | 0 | 0 | 0 | 0 | 0.617 |
| | TP | 36 | 0 | 6 | 0 | 0 | 0 | 0 | |
| | FP | 5 | 2 | 3 | 2 | 1 | 0 | 0 | |
| | TN | 6 | 346 | 238 | 349 | 355 | 340 | 355 | |
| | FN | 3 | 14 | 115 | 11 | 6 | 22 | 7 | |
| cVL: CLIP | F1 | 0.684 | 0.19 | 0 | 0 | 0 | 0 | 0.04 | 0.017 |
| | TP | 212 | 2 | 0 | 0 | 0 | 0 | 8 | |

| | | Polyp Detection | Adeno-carcinoma | Tubular adenoma | Tubulovillous adenoma | Villous adenoma | Hyperplastic polyp | Inflammatory polyp | Weighted F1 score |
|---|---|---|---|---|---|---|---|---|---|
| | FP | 196 | 2 | 0 | 3 | 0 | 8 | 385 | |
| | TN | 0 | 389 | 266 | 393 | 401 | 375 | 14 | |
| | FN | 0 | 15 | 142 | 12 | 7 | 25 | 1 | |
| cVL: BiomedCLIP | F1 | 0.887 | 0.563 | 0.293 | 0.167 | 0 | 0.211 | 0.043 | 0.277 |
| | TP | 188 | 9 | 29 | 4 | 0 | 6 | 3 | |
| | FP | 24 | 6 | 27 | 32 | 23 | 26 | 128 | |
| | TN | 172 | 385 | 239 | 364 | 378 | 357 | 271 | |
| | FN | 24 | 8 | 113 | 8 | 7 | 19 | 6 | |

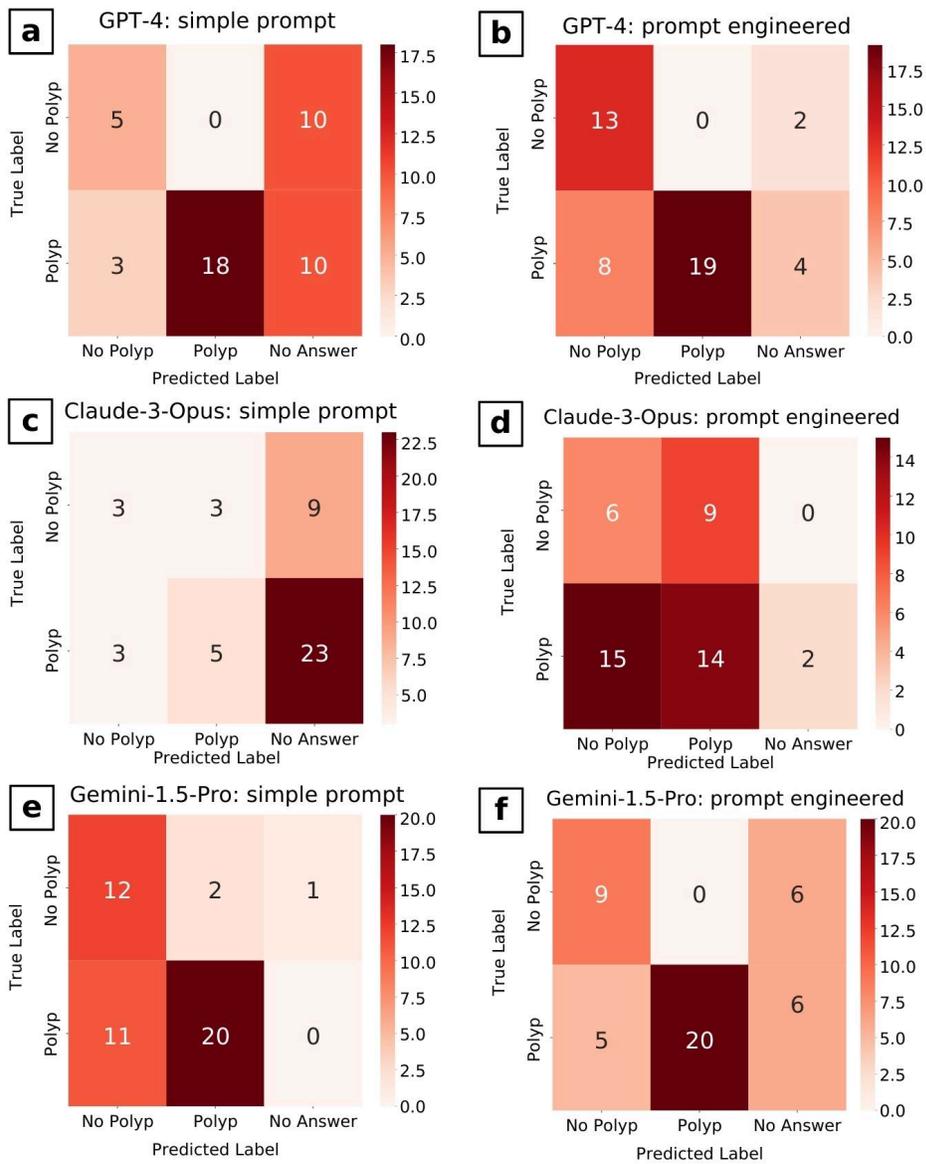

**Supplementary Figure S1.** The confusion matrix of polyp detection elucidates the results obtained from the simple prompt (left column: GPT-4 (a), Claude-3-Opus (c), Gemini-1.5-Pro (d)) alongside the outcomes produced from the engineered prompt (right column: GPT-4 (b), Claude-3-Opus (d), Gemini-1.5-Pro (f)).

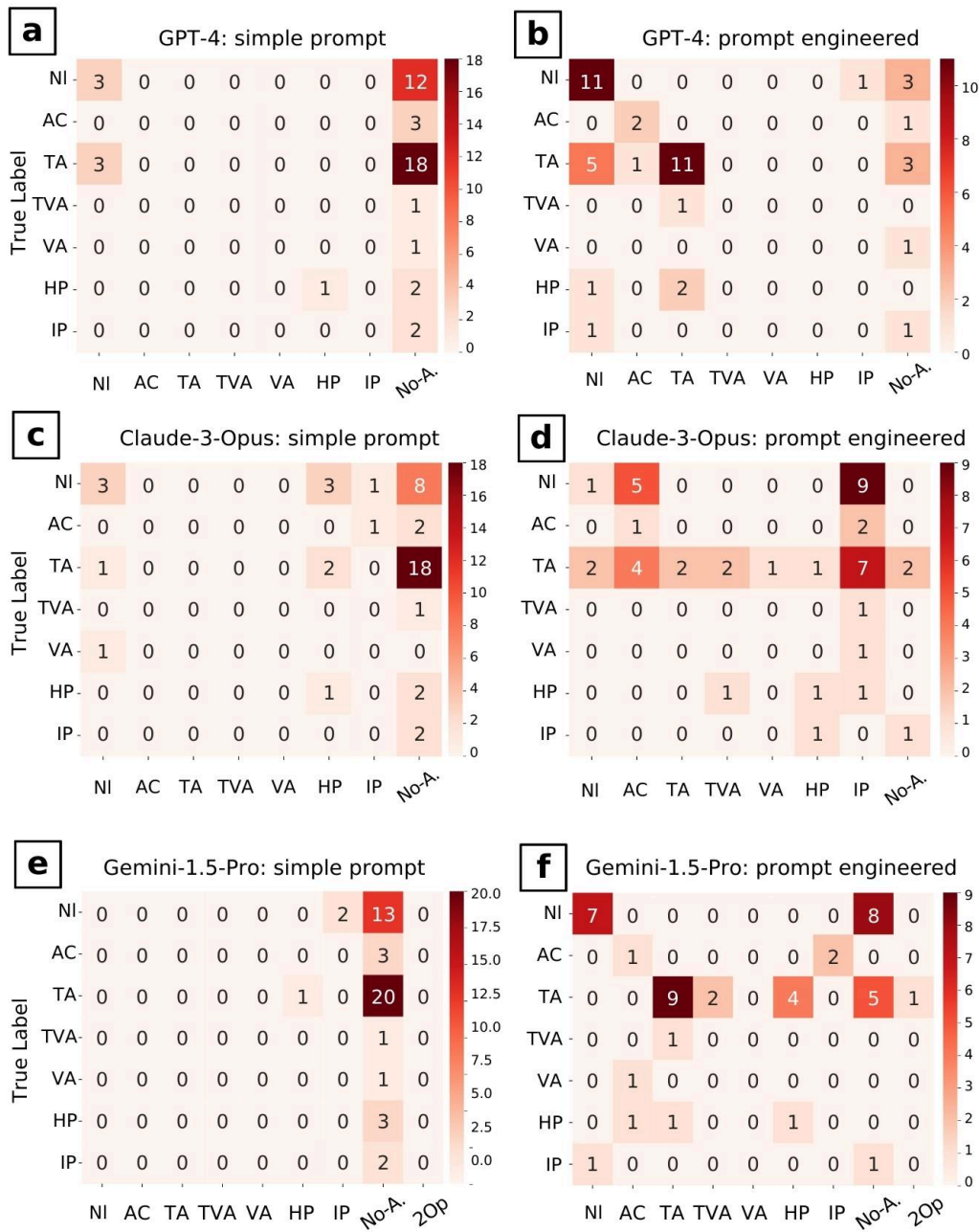

**Supplementary Figure S2.** The confusion matrix of polyp classification elucidates the results obtained from the simple prompt (left column: GPT-4 (a), Claude-3-Opus (c), Gemini-1.5-Pro (d)) alongside the outcomes produced from the engineered prompt (right column: GPT-4 (b), Claude-3-Opus (d), Gemini-1.5-Pro (f)). Abbreviations: AC, Adenocarcinoma; TA, Tubular Adenoma; TVA, Tubulovillous Adenoma; VA, Villous Adenoma; HP, Hyperplastic Polyp; IP, Inflammatory Polyp; No-A: No Answer provided; 2OP: Two options (polyp type) is selected.

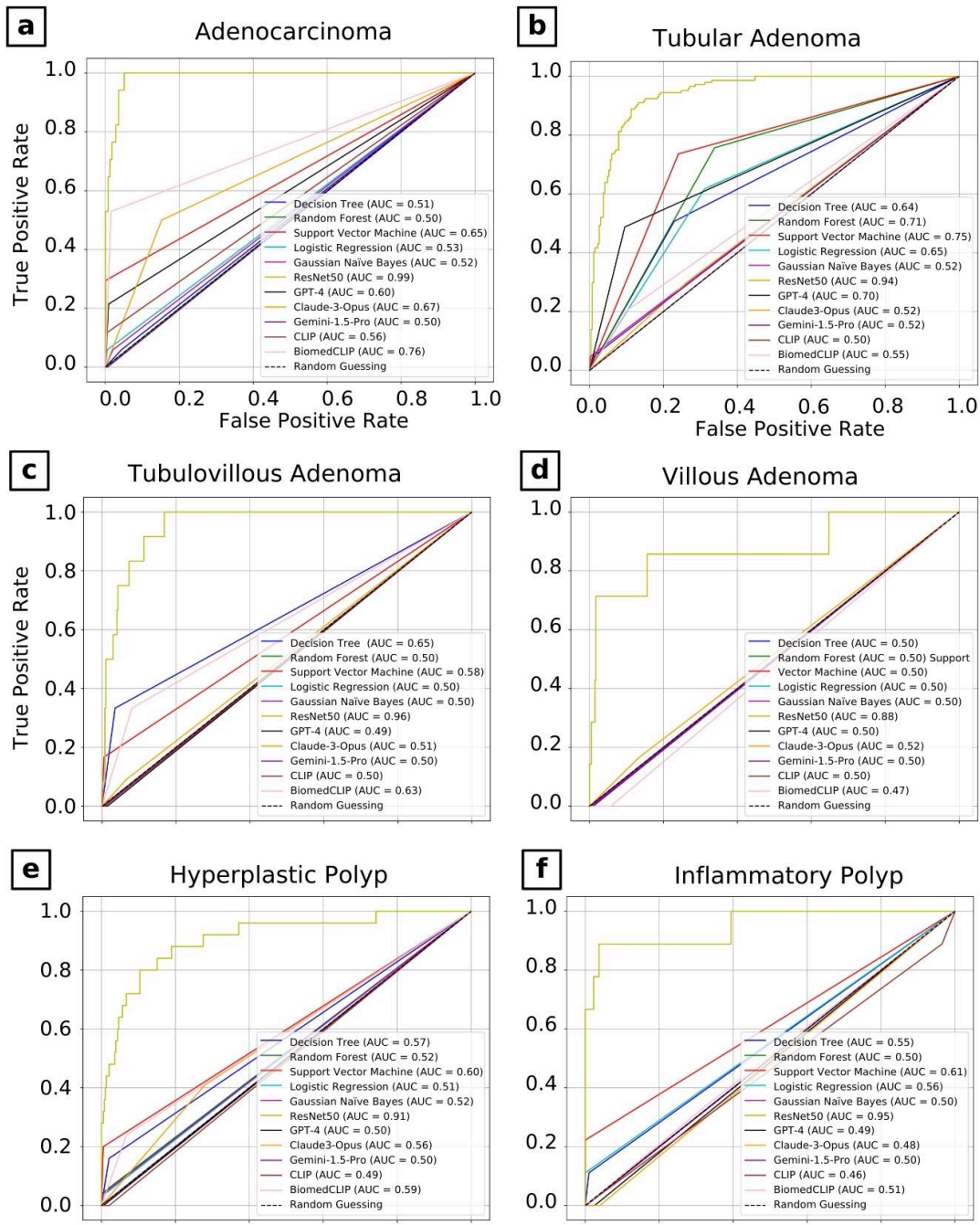

**Supplementary Figure S3.** The obtained receiver operating characteristic curve and the area under the curve (AUC) for the classification of polyp types, including adenocarcinoma (a), tubular adenoma (b), tubulovillous adenoma (c), villous adenoma (d), hyperplastic polyp (e), and inflammatory polyp (f).

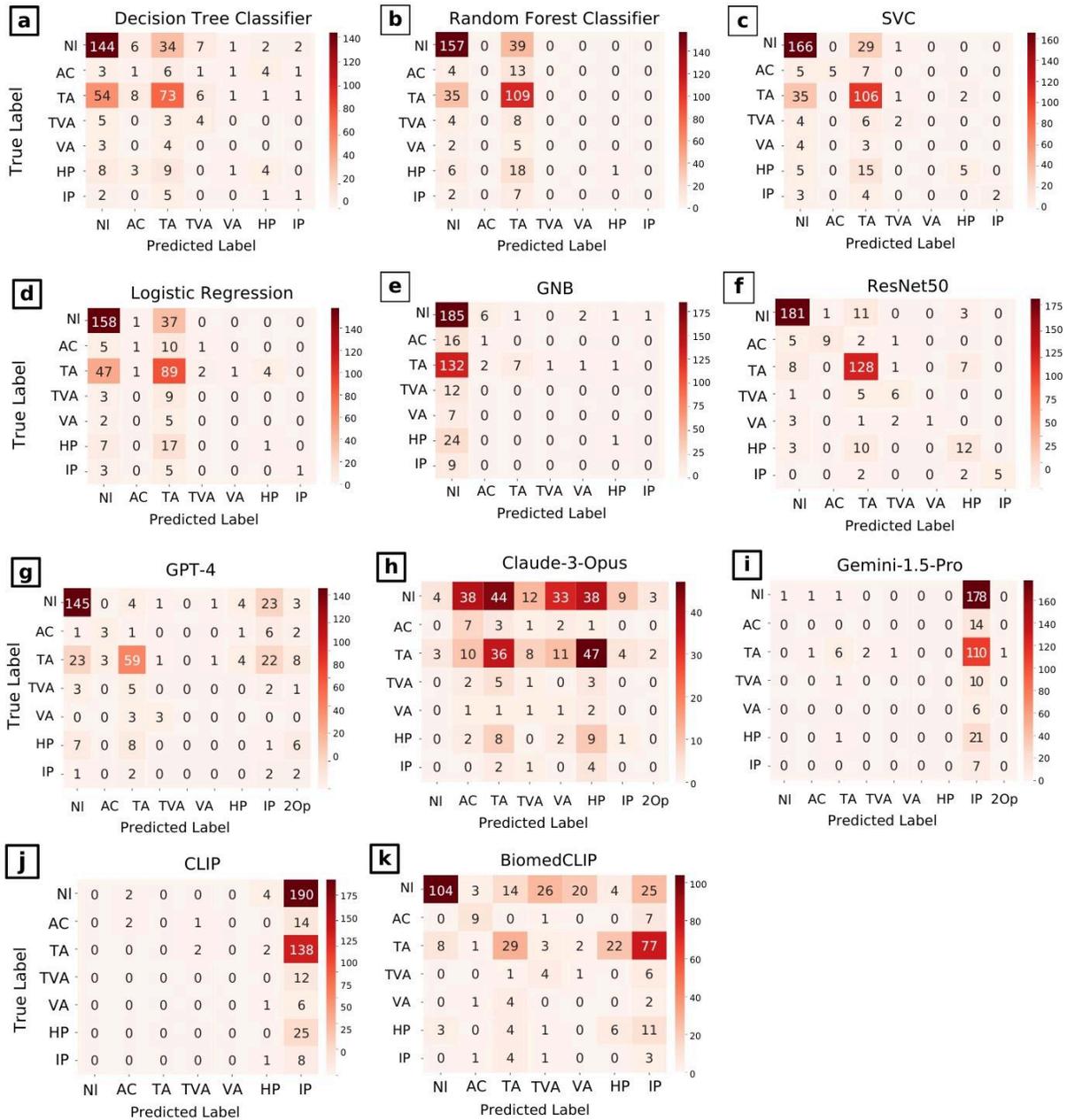

**Supplementary Figure S4.** Confusion matrices depicting polyp classification performance across various models: classical machine learning algorithms—Decision Tree (a), Random Forest (b), Support Vector Machine (c), Logistic Regression (d), Gaussian Naive Bayes (e); convolutional neural network—ResNet-50 (f); vision-language models—GPT-4 (g), Claude-3-Opus (h), Gemini-1.5-Pro (i); and contrastive vision-language encoders—CLIP (j), BiomedCLIP (k). Each matrix illustrates model predictions relative to ground-truth labels.